\documentclass[sigconf,natbib=true,nonacm]{acmart}
\usepackage{multirow}

\newcommand{\pith}{\pi_{\theta}}
\newcommand{\pisft}{\pi_{0}}
\newcommand{\piref}{\pi_{\mathrm{ref}}}
\newcommand{\KL}{\mathrm{KL}}
\newcommand{\cI}{\mathcal{I}}
\newcommand{\cR}{\mathcal{R}}
\newcommand{\cA}{\mathcal{A}}
\newcommand{\rank}{\mathrm{rank}}
\newcommand{\TopSet}{\mathrm{Top}}
\newcommand{\HR}{\mathrm{HR}}
\newcommand{\hit}[1]{hit@#1}
\newcommand{\ndcg}{NDCG@10}

\begin{document}

\title{Better Rankers, Worse Retrievers: The Hidden Cost of RL Post-Training in Generative Recommendation}

\author{Yuheng Chen}
\affiliation{%
  \institution{The Hong Kong University of Science and Technology}
  \city{Hong Kong}
  \country{China}}
\email{ychenlp@connect.ust.hk}

\author{Fan Zhang}
\affiliation{%
  \institution{The Hong Kong University of Science and Technology}
  \city{Hong Kong}
  \country{China}}

\author{Yixuan Xu}
\affiliation{%
  \institution{The Hong Kong University of Science and Technology}
  \city{Hong Kong}
  \country{China}}

\author{Chao Yang}
\affiliation{%
  \institution{Trip.com Group}
  \city{Shanghai}
  \country{China}}
\email{cyang3@trip.com}

\author{Yang Xiang}
\affiliation{%
  \institution{Trip.com Group}
  \city{Shanghai}
  \country{China}}
\email{yxiang@trip.com}

\author{Can Yang}
\authornote{Corresponding author.}
\affiliation{%
  \institution{The Hong Kong University of Science and Technology}
  \city{Hong Kong}
  \country{China}}
\email{macyang@ust.hk}

\begin{abstract}
Generative recommenders decode the semantic ID of a user's next item with a language model, and post-training them with reinforcement learning (RL) after supervised fine-tuning (SFT) has become standard practice.
The gains are mostly reported on the first ten results, yet in deployment such models hand hundreds of candidates to a downstream ranker or reward model.
We show that RL post-training has a hidden cost: it improves accuracy at the top of the list (\hit{$K$} for small $K$) but, on all five datasets we study, lowers the recall of the candidate list (\hit{$K$} for $K$ in the hundreds).
The model becomes a better ranker and a worse retriever.
Controlled experiments trace the loss to the negative advantages that RL assigns to the candidates it samples, the rollout.
The root cause is that only one next item is logged per user, so RL treats every other candidate as wrong.
Yet these candidates are the model's most plausible alternatives, and often the next items of other users with similar histories.
Penalising them pushes down exactly what similar users choose next, and these items drop out of those users' candidate lists.
This diagnosis points directly to a remedy: out-of-rollout anchoring, a single additional loss term that preserves the SFT model's ordering of the items outside the rollout.
It keeps the gain at the top and restores recall to the SFT level on all five datasets, and it is effective under other RL configurations as well.
To our knowledge, this is the first systematic study of this recall loss: we measure it, identify its cause, and show that, once the cause is known, a simple remedy suffices.
\end{abstract}

\ccsdesc[500]{Information systems~Recommender systems}
\ccsdesc[500]{Information systems~Retrieval models and ranking}
\ccsdesc[300]{Computing methodologies~Reinforcement learning}

\keywords{generative recommendation, semantic IDs, reinforcement learning, GRPO, retrieval, recall}

\maketitle

\section{Introduction}\label{sec:intro}

Generative recommendation represents each item by a short sequence of discrete tokens, its \emph{semantic ID} (SID), and trains a sequence model to generate the SID of a user's next item~\citep{rajput2023tiger}.
With language models as backbones~\citep{zheng2024lcrec,he2025plum,kong2025minionerec}, training has settled into two stages: supervised fine-tuning (SFT) on logged next items, followed by post-training with reinforcement learning (RL), most often GRPO~\citep{shao2024deepseekmath}.
This pipeline has since been adopted widely, in production systems and in open research alike~\citep{deng2025onerec,zhou2025onerectr,zhou2025onerecv2,kong2025minionerec,liu2025onerecthink,zhou2025openonerec,yu2026daso,zheng2026sapo,chen2026lasar}.
In the RL stage the policy generates a few candidate items for a user (a \emph{rollout}), the candidate that matches the logged next item is rewarded, and the benefit is mostly reported as a higher hit rate or NDCG at small $K$, typically within the first ten results.

\begin{figure}[t]
  \centering
  \includegraphics[width=.93\linewidth]{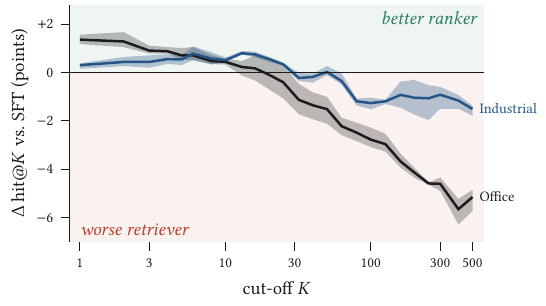}
  \caption{Change in \hit{$K$} after GRPO post-training on Office and Industrial, relative to the SFT model that RL started from (exact ranking of the full catalogue; mean of three seeds, bands show their range).}
  \Description{Line chart of the change in hit at K against the cut-off K on a logarithmic axis for Office and Industrial. Both curves are above zero for small K and fall below zero beyond K of about 20 and 50, reaching minus 5.1 and minus 1.5 points at K equal to 500.}
  \label{fig:phenomenon}
\end{figure}

\begin{figure*}[t]
  \centering
  \includegraphics[width=\linewidth]{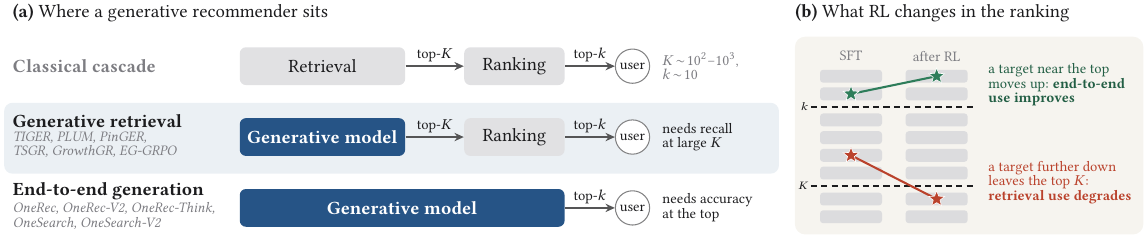}
  \caption{(a) A generative recommender is deployed as the retrieval stage of a cascade (highlighted: the common case), where recall at large $K$ matters, or end to end, where accuracy at the top matters (example systems by the role their papers describe).
  (b) The effect studied here: after RL post-training, targets near the top move up and targets further down drop out of the top $K$.}
  \Description{Two panels. Panel a shows the classical retrieval and ranking cascade, a generative model replacing the retrieval stage, which is highlighted as the common case, and a generative model replacing the whole cascade. Panel b shows two rankings side by side, under SFT and after RL: a target near the top moves up and a target further down leaves the top K.}
  \label{fig:roles}
\end{figure*}

How much of the generated list is used depends on the deployment (Figure~\ref{fig:roles}a, Table~\ref{tab:prior}).
In \emph{generative retrieval}, the common case, the model is one retrieval channel: it hands its top $K$ candidates, with $K$ from a hundred to more than a thousand, to downstream rankers~\citep{rajput2023tiger,he2025plum,li2026pinger,zhan2026tsgr}, and it is judged by whether the user's next item is among them.
Such models are often post-trained with RL (Table~\ref{tab:prior}), but its effect is seldom measured at these depths.
For example, benchmark studies stop at $K=32$ or below, GenRec at $K=50$~\citep{zou2026genrec} and EG-GRPO at ten~\citep{zhu2026eggrpo}.
Scattered hints of a problem do appear in these papers: TSGR, which passes 1{,}500 candidates to its ranker, finds the gain from RL marginal and has not deployed it~\citep{zhan2026tsgr}, and EG-GRPO reports that standard GRPO slightly raises the click hit rate of the top result but lowers it already within the top ten~\citep{zhu2026eggrpo}.
In \emph{end-to-end generation}, pioneered by OneRec~\citep{deng2025onerec}, the model replaces the retrieve-then-rank cascade, and its first few outputs are shown to the user.
Yet even OneRec is at its best as a candidate generator: it produces 512 candidates per request, and in its online test showing the model's own top results merely matches the production cascade ($+0.01\%$ app stay time), whereas letting a reward model select among the candidates brings $+0.54\%$~\citep{zhou2025onerectr}.
In its best-performing variant, the generative model is thus a retrieval stage again.

In both roles, then, the list is used well beyond its top, yet that is where the effect of RL is usually reported: OneSearch-V2, for instance, decodes with a beam of 512 and compares its models with and without RL on the first ten results~\citep{chen2026onesearchv2}.
We therefore ask three questions.
\textbf{RQ1}: What does RL post-training do to the ranking beyond the top of the list?
\textbf{RQ2}: What causes the change, and why does it arise in recommendation?
\textbf{RQ3}: Can it be avoided without giving up the gain at the top?

\begin{table}[t]
  \caption{RL and preference post-training of generative recommenders in prior work.
  Candidates: number generated per request (passed to the next stage or, failing that, the largest beam width); n/s: not stated.
  $^{\dagger}$Online results are reported with and without a reward model that selects among the generated candidates.}
  \label{tab:prior}
  \small
\begin{tabular}{@{}l@{\hspace{10pt}}l@{\hspace{10pt}}l@{\hspace{10pt}}r@{}}
\toprule
System & Platform & Objective & Candidates \\
\midrule
\multicolumn{4}{@{}l}{\textit{Retrieval stage: candidates are passed to a ranker}} \\
TSGR~\citep{zhan2026tsgr} & Taobao & GRPO & 1{,}500 \\
GrowthGR~\citep{wang2026growthgr} & Taobao & GRPO & 1{,}000 \\
EG-GRPO~\citep{zhu2026eggrpo} & Tmall & GRPO & 100 \\
GenR-PO~\citep{li2024genrpo} & JD & DPO & 100 \\
PinGER~\citep{li2026pinger} & Pinterest & DPO & 100 \\
GenRec~\citep{zou2026genrec} & JD & GRPO & n/s \\
\addlinespace[2pt]
\multicolumn{4}{@{}l}{\textit{End-to-end generation, also run as retrieval for a reward model}} \\
OneRec$^{\dagger}$~\citep{deng2025onerec,zhou2025onerectr} & Kuaishou & DPO, ECPO & 512 \\
OneRec-V2~\citep{zhou2025onerecv2} & Kuaishou & GBPO & 512 \\
OneSearch$^{\dagger}$~\citep{chen2025onesearch} & Kuaishou & DPO & 512 \\
OneSearch-V2~\citep{chen2026onesearchv2} & Kuaishou & GRPO & 512 \\
\addlinespace[2pt]
\multicolumn{4}{@{}l}{\textit{Benchmark studies}} \\
MiniOneRec~\citep{kong2025minionerec} & -- & GRPO & -- \\
OpenOneRec~\citep{zhou2025openonerec} & -- & GRPO & -- \\
Rank-GRPO~\citep{zhu2025rankgrpo} & -- & GRPO & -- \\
\bottomrule
\end{tabular}
\end{table}

\textbf{RQ1.}
We score every catalogue item for every test user, with the SFT model and with the RL models trained from it, and compare \hit{$K$} up to $K=500$ in paired tests, on five datasets with three RL seeds.
Under the most common configuration (GRPO with a hit-based reward on beam rollouts; Section~\ref{sec:background}), RL improves the top of the list, as reported, and lowers \hit{$K$} beyond the first few dozen results: the gain turns into a loss at $K\approx 20$ on Office and $K\approx 50$ on Industrial (Figure~\ref{fig:phenomenon}).
On Office, \ndcg{} rises from 11.59\% to 12.55\%, while \hit{50} falls from 25.73\% to 24.22\% and \hit{500} from 52.84\% to 47.69\%; 22\% of the test targets in the SFT model's top 500 are no longer there.
Judged by its first few results, the model is a better ranker; judged by the candidates it passes on, it is a worse retriever.
The loss appears on all five datasets, widely used benchmarks that differ in domain and catalogue size; it persists under beam search, and an evaluation confined to the top of the list sees only the improvement.

\textbf{RQ2.}
The objective is not to blame: the exact solution of the KL-regularised RL problem cannot lower the expected \hit{$K$} at any $K$ (Proposition~\ref{prop:ideal}).
The loss comes from how the update is estimated.
SFT raises the likelihood of the logged item and names no item as wrong.
GRPO also lowers every sampled candidate that is not the logged item.
In the reasoning tasks it was designed for, such an answer is wrong; in recommendation it is among the policy's top choices for the user, a plausible next item that happened not to be the one recorded, akin to the unlabelled positives of dense retrieval (Section~\ref{sec:whyrec}).
All users share one small item set, so such an item is penalised for many users, far more often than it is rewarded, and its score falls for users with similar histories too, including those whose next item it is.
If it lies outside such a user's rollout, the rollout earns no reward and nothing pushes back, so rarely sampled items move above it (Figure~\ref{fig:causes}).
Controlled experiments support this account.
Removing the negative advantages removes the loss, and the gain with it; keeping only them makes the loss at least three times larger.
And the loss falls where the penalties fall: the test users who lose the most recall are those whose next item is penalised most, and these items can be identified from the SFT model before RL starts.

\textbf{RQ3.}
The cause points to the remedy: if the damage is done outside the rollout, protecting the items there should be enough.
\emph{Out-of-rollout anchoring} does exactly this: it leaves GRPO unchanged inside the rollout and adds, for every prompt, a term that keeps the relative order the SFT model gave to $M$ items sampled outside the rollout (Figure~\ref{fig:anchor}).
On Office the anchor recovers 4.4 of the 5.1 points of \hit{500} lost under GRPO and keeps the gain at the top; on the other four datasets \hit{500} returns to the level of the SFT model or slightly above it, and \ndcg{} matches that of GRPO on all five.
The recovery starts as early as the loss: at $K=50$ the anchored model is the best of the three on all five datasets.
The recall comes back where it was lost: for the test users whose next item is penalised most, the anchor raises \hit{500} by 20 points over GRPO on Office.
The anchor has the same effect on other RL configurations.

In summary, our main contribution is a diagnosis, and the remedy follows from it.
We contribute
(i) a measurement, by exact full-catalogue ranking on five datasets, of a recall loss under RL post-training that evaluation at the top of the list hides;
(ii) an explanation that traces the loss to negative advantages on plausible candidates, which accumulate per item and carry over to similar users, and that shows why the problem arises in recommendation;
(iii) controlled experiments that support it, including a record computed before RL that identifies the users who lose the most recall; and
(iv) out-of-rollout anchoring, a one-term remedy derived from this explanation, which recovers the lost recall and keeps the gain at the top.
\section{Preliminaries and Setup}\label{sec:background}

\subsection{Generative recommendation with semantic IDs}
Let $\cI$ be the item catalogue.
Each item $i$ has a semantic ID $c(i)=(c_1,c_2,c_3)$ of three codes from codebooks of size 256, obtained by residual quantisation of a text embedding of the item~\citep{lee2022rqvae,rajput2023tiger}; items with similar content share their leading codes.
The codes are added to the vocabulary of a language model.
Given a prompt $x$ that lists the SIDs of the user's recent items, the model scores an item by generating its codes in order,
\begin{equation}
  s_\theta(i\mid x)=\log\pith(i\mid x)=\sum_{l=1}^{3}\log\pith\bigl(c_l(i)\mid x,\,c_{<l}(i)\bigr).
  \label{eq:score}
\end{equation}
Decoding is constrained to the prefix tree of valid SIDs~\citep{decao2021genre}, and a beam search of width $B$ returns approximately the $B$ highest-scoring items.
As a retrieval channel the model returns several hundred items to a ranker; in end-to-end use about ten are shown (Figure~\ref{fig:roles}a).
We write $K$ for a generic cut-off.

\subsection{SFT and RL post-training}
SFT maximises $\log\pith(y\mid x)$ over logged pairs of a history $x$ and its next item $y$, the \emph{target}, together with auxiliary tasks that align SIDs with item titles~\citep{kong2025minionerec}.
We call the result the SFT model $\pisft$.

For RL we follow MiniOneRec's instantiation of GRPO~\citep{kong2025minionerec,shao2024deepseekmath}; other SID-based pipelines share its structure.
For each prompt the policy produces a \emph{rollout} $\cR(x)=(i_1,\dots,i_G)$, the group of GRPO, of $G=16$ distinct candidates by constrained beam search with sampling, which concentrates on the items the policy ranks highest.
With $\rho(j)$ the position of candidate $i_j$ in the beam, the reward is
\begin{equation}
  r_j=\begin{cases}
    1 & \text{if } i_j=y,\\
    -w_{\rho(j)} & \text{if } i_j\neq y \text{ and } y\in\cR(x),\\
    0 & \text{if } y\notin\cR(x),
  \end{cases}
  \quad w_\rho\propto\frac{1}{\log_2(\rho+1)},
  \label{eq:reward}
\end{equation}
with the weights $w_\rho$ normalised to sum to one over the $G$ positions.
The rank-aware penalty is larger for candidates the policy ranks higher; MiniOneRec introduces it to push the model ``away from hard negatives''.
A rollout that does not contain the target earns zero reward for every candidate, as in the released code.
Rewards are standardised within the group, $A_j=(r_j-\bar r)/(\sigma_r+\epsilon)$: when the target is in the rollout it receives a positive advantage and each of the other $G-1$ candidates a negative one; otherwise all advantages are zero.
The loss for one prompt is
\begin{equation}
  \mathcal{L}_{\mathrm{GRPO}}(\theta;x)=-\frac{1}{G}\sum_{j=1}^{G}A_j\,\log\pith(i_j\mid x)+\beta\,\widehat{\KL}_{\cR(x)}\bigl(\pith\,\|\,\piref\bigr),
  \label{eq:grpo}
\end{equation}
where the second term is the reverse KL divergence to a reference policy $\piref$, estimated on the tokens of the rollout.
The implementation uses the clipped surrogate of PPO~\citep{schulman2017ppo,vonwerra2020trl}; each rollout is used for one gradient step, so the importance ratio equals one and the gradient is that of Eq.~\eqref{eq:grpo}.
For the same reason, variants that redefine the ratio or its clipping, such as GSPO, DAPO, CISPO and ECPO~\citep{zheng2025gspo,yu2025dapo,minimax2025m1,zhou2025onerectr}, have the same policy-gradient term in this setting.

\paragraph{Configuration}
Published SID-based pipelines that retain a KL term use $\beta=10^{-3}$~\citep{liu2025onerecthink,yu2026daso,zheng2026sapo,chen2026lasar} and, where they state it, a fixed reference, the SFT checkpoint or that of the preceding stage.
We adopt it, with $\piref=\pisft$, and refer to it simply as GRPO.

\subsection{Evaluation}
For each test prompt we compute $s_\theta(i\mid x)$ for every item by expanding the SID prefix tree and record the exact rank of the target,
$\rank_\theta(y\mid x)=1+|\{i\in\cI: s_\theta(i\mid x)>s_\theta(y\mid x)\}|$.
We report \hit{$K$}, the share of test prompts with $\rank_\theta(y\mid x)\le K$, for $K$ up to 500 (14\% of the smallest catalogue), and \ndcg{}~\citep{jarvelin2002cumulated}.
Each test prompt has one target, so \hit{$K$} equals recall at $K$, and we speak of \emph{recall} for \hit{$K$} at the cut-offs of a retrieval stage.
Exact ranking removes the approximation of beam search and makes every $K$ observable in one pass; as a check, a beam search run on our scores reproduces about 98\% of the top-10 results of MiniOneRec's evaluation script.
All comparisons are paired: two models are evaluated on the same prompts, and a difference counts as significant if its 95\% bootstrap interval over prompts excludes zero~\citep{efron1994bootstrap}.
The main comparisons use three RL runs with different seeds; we report their mean.
For single runs a star marks a significant difference.

\subsection{Experimental setup}\label{sec:setup}
\begin{table}[t]
  \caption{Datasets. ``In rollout'': share of training prompts whose rollout contains the target at the start of RL.}
  \label{tab:data}
  \small
  \begin{tabular}{@{}lrrlr@{}}
    \toprule
    Dataset & Items & Test & Split & In rollout \\
    \midrule
    Office     & 3{,}459  & 4{,}866 & chronological & 37\% \\
    Industrial & 3{,}686  & 4{,}533 & chronological & 42\% \\
    Beauty     & 12{,}101 & 5{,}000 & leave-one-out & 24\% \\
    Toys       & 11{,}924 & 5{,}000 & leave-one-out & 16\% \\
    Sports     & 18{,}357 & 5{,}000 & leave-one-out & 11\% \\
    \bottomrule
  \end{tabular}
\end{table}
Table~\ref{tab:data} lists the datasets.
Office and Industrial are the benchmarks released with MiniOneRec, built from the 2018 Amazon reviews~\citep{ni2019justifying} with a chronological 8:1:1 split; we use their files and SIDs unchanged.
Beauty, Toys and Sports are the 5-core subsets of the 2014 Amazon reviews~\citep{mcauley2015image,he2016ups} with the usual leave-one-out split~\citep{kang2018sasrec,rajput2023tiger}; we build their SIDs with the MiniOneRec pipeline and evaluate on 5{,}000 sampled users.
The backbone is Qwen2.5-1.5B-Instruct~\citep{qwen2024qwen25}, fully fine-tuned.
SFT follows the MiniOneRec recipe with early stopping on validation loss.
RL uses $G=16$, a learning rate of $10^{-5}$, 32 prompts per step and gradient clipping at 0.3.
These are the settings of the MiniOneRec code, which was developed on Office and Industrial; we apply them to every dataset without tuning them for the three datasets we add.
As in that code, about 70\% of the RL prompts ask for the next item given the SID history, and the rest are two auxiliary tasks with item titles in the input.
One epoch is about 1{,}700 steps on every dataset (on Beauty, Toys and Sports, RL draws on a subset of the training sequences as large as the Office training set), and every model is evaluated after one epoch.
On each dataset all RL runs start from the same SFT checkpoint; the seed changes the order of the training prompts and the sampling of rollouts and anchor items.
\section{RL Post-Training Trades Recall for Top-of-List Accuracy}\label{sec:phenomenon}

\subsection{Gains at the top, losses below}

\begin{table}[t]
  \caption{Change after one epoch of GRPO relative to SFT, in points (mean of three seeds). Table~\ref{tab:main} gives the absolute values.}
  \label{tab:phenomenon}
  \small\setlength{\tabcolsep}{4.2pt}
\begin{tabular}{@{}l@{\hspace{6pt}}r@{\hspace{11pt}}rrrrrr@{\hspace{5pt}}}
\toprule
 & NDCG & \multicolumn{6}{c}{hit@$K$} \\
\cmidrule(l){3-8}
 & @10 & 5 & 10 & 50 & 100 & 200 & 500 \\
\midrule
Office & ${+}0.95$ & ${+}0.71$ & ${+}0.44$ & ${-}1.52$ & ${-}2.77$ & ${-}4.14$ & ${-}5.14$ \\
Industrial & ${+}0.43$ & ${+}0.54$ & ${+}0.51$ & ${+}0.02$ & ${-}1.27$ & ${-}1.04$ & ${-}1.51$ \\
Beauty & ${+}0.38$ & ${+}0.28$ & ${+}0.47$ & ${+}0.33$ & ${-}0.31$ & ${-}0.39$ & ${-}0.49$ \\
Toys & ${+}0.40$ & ${+}0.30$ & ${+}0.50$ & ${+}0.24$ & ${-}0.40$ & ${-}0.62$ & ${-}0.04$ \\
Sports & ${+}0.01$ & ${+}0.20$ & ${-}0.18$ & ${-}0.47$ & ${-}0.25$ & ${-}0.19$ & ${-}0.67$ \\
\bottomrule
\end{tabular}
\end{table}

Table~\ref{tab:phenomenon} reports the change after one epoch of GRPO, and Figure~\ref{fig:phenomenon} shows it at every $K$ for Office and Industrial~\citep{kong2025minionerec}.
At the top of the list RL helps, as prior work reports: on four of the five datasets \ndcg{} and \hit{$K$} rise throughout the first ten results, and on Sports \hit{$K$} rises within the first five.
Further down it hurts, and it does not take a deep list: the gain turns into a loss at $K\approx 20$ on Office and $K\approx 50$ on Industrial.
On Office the loss is significant in every seed from $K=50$ upwards (1.4\% of the catalogue) and reaches 5.1 points at $K=500$; on Industrial it is significant at $K=100$ and $K=500$ in every seed and amounts to 0.9 to 1.5 points for $K\ge 100$.
On Beauty, Toys and Sports, whose catalogues are three to five times larger and whose SFT models are weaker, \hit{$K$} is lower than under SFT at every $K\ge 100$, by up to 0.8 points.
We did not tune the RL hyper-parameters for these datasets (Section~\ref{sec:setup}), so RL is probably not at its best there; the trend is nonetheless the same on all five datasets.
The reordering behind these numbers is also of similar size everywhere: between 14\% and 22\% of the test targets that the SFT model ranked in its top 500 are no longer there after RL.
The loss falls on the part of the list that deployed systems use.

The loss is also there when the candidates are produced by beam search, as they are in deployed systems.
With a constrained beam of width $B=K$, \hit{200} changes by $-2.59$ points on Office and $-1.96$ on Industrial, and \hit{500} by $-3.06$ and $-1.83$.
Under the standard protocol, a beam of 50 scored at $K=10$, \hit{10} changes by $+0.39$ and $+0.46$ points: the protocol sees the gain and none of the loss.

The effect is not tied to this configuration (Table~\ref{tab:pairs}, rows without the anchor).\label{sec:robust}
With a synced reference, one that is moved towards the policy during training~\citep{kong2025minionerec}, GRPO loses 6.4 points of \hit{500} on Office and 1.5 on Industrial, and a second epoch deepens these losses to 8.9 and 2.1 points.
REINFORCE++ with baseline~\citep{hu2025reinforcepp}, a second policy-gradient algorithm, with advantages normalised over the batch, loses three to four times as much recall as GRPO.

\begin{figure*}[t]
  \centering
  \includegraphics[width=.95\linewidth]{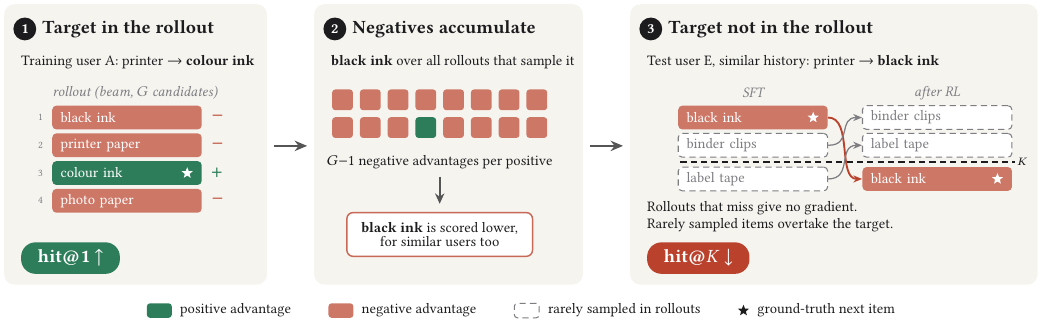}
  \caption{Why recall degrades (illustrative example).
  (1)~A target in the rollout gets a positive advantage and the other candidates negative ones, which improves the top of the list.
  (2)~A plausible item such as black ink is sampled for many users and is rarely the target, so penalties accumulate on it and its score also falls for similar users.
  (3)~Where the rollout misses black ink no gradient counteracts the fall, and for a test user whose target it is, rarely sampled items overtake it.}
  \Description{Three cards. Card one shows a rollout for a training user in which the target, colour ink, gets a plus sign and three other candidates including black ink get minus signs. Card two shows sixteen squares of which fifteen are negative, standing for the advantages black ink collects over all rollouts that sample it. Card three shows a test user whose target is black ink: under SFT it is above the top K line, after RL two rarely sampled items have moved above it and it falls below the line.}
  \label{fig:causes}
\end{figure*}

\subsection{Mid-ranked items lose rank}

\begin{table}[t]
  \caption{Where the targets move under GRPO: test prompts grouped by the rank of their target under SFT, with the median rank before and after RL and the share whose rank worsens (mean of three seeds).}
  \label{tab:bins}
  \small
\begin{tabular}{@{}l@{\hspace{8pt}}rr@{\hspace{10pt}}rr@{}}
\toprule
 & \multicolumn{2}{c}{Office} & \multicolumn{2}{c}{Industrial} \\
\cmidrule(lr){2-3}\cmidrule(l){4-5}
Rank under SFT & median rank & demoted & median rank & demoted \\
\midrule
1 & 1 $\rightarrow$ 1 & 13\% & 1 $\rightarrow$ 1 & 19\% \\
2--10 & 4 $\rightarrow$ 3 & 42\% & 5 $\rightarrow$ 5 & 47\% \\
11--50 & 26 $\rightarrow$ 44 & 62\% & 25 $\rightarrow$ 30 & 56\% \\
51--100 & 72 $\rightarrow$ 136 & 64\% & 72 $\rightarrow$ 90 & 57\% \\
101--200 & 146 $\rightarrow$ 262 & 65\% & 147 $\rightarrow$ 166 & 55\% \\
201--500 & 326 $\rightarrow$ 497 & 63\% & 336 $\rightarrow$ 355 & 54\% \\
$>$500 & 1573 $\rightarrow$ 1563 & 44\% & 1590 $\rightarrow$ 1556 & 43\% \\
\bottomrule
\end{tabular}
\end{table}

Table~\ref{tab:bins} groups the test prompts by the rank the SFT model gave their target.
Targets ranked first mostly stay first, and targets at ranks 2 to 10 move up and down in about equal proportion.
On Office, targets at ranks 11 to 500 are demoted nearly twice as often as promoted and their median rank grows by half or more (on Industrial 54 to 57\% are demoted); targets beyond rank 500 improve slightly.
RL thus damages the middle of the ranking: the items that the SFT model considered plausible for a user without placing them at the very top.

This is a change of order, which sharpening the distribution cannot produce.
Rank-based metrics are invariant to monotone transformations of the scores, such as a lower temperature.
RL does make the distribution more peaked (on Office the entropy of the item distribution, averaged over test prompts, falls from 6.0 to 3.1 nats), but that alone would leave \hit{$K$} unchanged at every $K$: the loss requires pairs of items that the SFT model ordered correctly and the RL model inverts.
\section{Why Recall Degrades}\label{sec:causes}

Figure~\ref{fig:causes} summarises our answer to RQ2.
We show that the RL objective is not responsible (Section~\ref{sec:ideal}), locate the problem in the way it is estimated (\ref{sec:update}), test this account in controlled experiments (\ref{sec:evidence}) and explain why it arises in recommendation (\ref{sec:whyrec}).

\subsection{The ideal objective is not the cause}\label{sec:ideal}
Let $p(i\mid x)$ be the probability that item $i$ is the next item after history $x$; the logged target is one draw $y\sim p(\cdot\mid x)$.
The expected hit rate of the top-$K$ list of a model $\pi$ is $\HR_K(\pi;x)=\sum_{i\in\TopSet_K(\pi(\cdot\mid x))}p(i\mid x)$.
Under the hit reward $\mathbf{1}[i=y]$, a sampled item $i$ earns $p(i\mid x)$ in expectation, so RL with a KL penalty to $\pisft$ maximises
$J(\pi;x)=\sum_{i}\pi(i\mid x)\,p(i\mid x)-\beta\,\KL(\pi(\cdot\mid x)\,\|\,\pisft(\cdot\mid x))$,
whose maximiser is the exponential tilt $\pi^\star(i\mid x)\propto\pisft(i\mid x)\exp(p(i\mid x)/\beta)$~\citep{korbak2022rlkl}.

\begin{proposition}\label{prop:ideal}
For every history $x$, every $\beta>0$ and every $K$, $\HR_K(\pi^\star;x)\ge\HR_K(\pisft;x)$.
\end{proposition}
\begin{proof}
For two items $a$ and $b$,
$\pi^\star(a\mid x)/\pi^\star(b\mid x)=[\pisft(a\mid x)/\pisft(b\mid x)]\exp((p(a\mid x)-p(b\mid x))/\beta)$.
So if $\pisft$ ranks $b$ above $a$ and $\pi^\star$ ranks $a$ above $b$, then $p(a\mid x)>p(b\mid x)$.
Hence every item that enters the top $K$ under $\pi^\star$ has a larger $p$ than every item that leaves it.
Equally many items enter and leave, so the sum of $p$ over the top $K$ does not decrease.
\end{proof}

An exactly optimised objective would thus sharpen the distribution and change the order of a pair of items only in the direction of their true probabilities.
The loss we observe therefore consists of new inversions, introduced by three departures from this ideal: the reward is a single logged item instead of $p$, the update touches only the sampled candidates, and one network serves all users.

\subsection{What the update does}\label{sec:update}
For one prompt, the gradient of the first term of Eq.~\eqref{eq:grpo} is, up to sign and writing $\cR$ for $\cR(x)$,
\begin{equation}
  g(x)=\frac{\mathbf{1}[y\in\cR]}{G}\Bigl(A_y\nabla\log\pith(y\mid x)-\!\!\!\sum_{i\in\cR\setminus\{y\}}\!\!\!|A_i|\nabla\log\pith(i\mid x)\Bigr).
  \label{eq:update}
\end{equation}
The SFT gradient is $\nabla\log\pith(y\mid x)$: every prompt contributes, and no item is singled out to be lowered.
Eq.~\eqref{eq:update} differs in two ways, and generalisation turns the difference into a recall loss.

\paragraph{It names negatives}
Whenever the target is in the rollout, the other $G-1$ candidates are pushed down.
This improves the top of the list, because the target rises above the items that competed with it (step~1 in Figure~\ref{fig:causes}).
But those candidates are among the policy's highest-ranked alternatives for this user, and with one logged next item per history such an alternative is not thereby a wrong recommendation.
Over training, an item collects a negative advantage each time it is sampled next to another user's target and a positive one only when it is itself the target (step~2).
The items penalised in this way are, to a large extent, other users' answers.
To measure this before any RL, we take the SFT model's 16 highest-scoring candidates for every training prompt, compute their advantages and sum them per item; we call the sum the item's \emph{rollout record}.
It is negative for the target of 45\% of the Office test prompts and 34\% of the Industrial ones, and only 1.7\% of the test targets never appear among these candidates.

\paragraph{It is gated}
A prompt contributes only if its rollout contains the target, which at the start of RL holds for about 40\% of the training prompts on Office and Industrial and for 11 to 24\% on the other datasets (Table~\ref{tab:data}).
For all other prompts the policy-gradient term is silent: if their target has been lowered by updates that came from other prompts, nothing in their own update restores it (step~3).

\paragraph{Generalisation spreads the penalties}
Consider a tabular policy in which every history has its own softmax over items with free logits.
A step along $g(x)$ would change only the logits of the $G$ sampled items of $x$, because the advantages of a group sum to zero: items outside the rollout would keep their mutual order, and a target outside its own rollout could never be lowered.
Any change in the rank of an out-of-rollout target therefore comes from generalisation across histories, since the SID tokens and the network are shared by all users.
A recommender is useful because it generalises in this way, and the same generalisation carries a penalty from the users who sampled an item to users with similar histories, including those for whom the item is the right answer.
Items that are rarely sampled for anyone receive no penalties, keep their scores, and overtake the penalised one.

\subsection{Evidence}\label{sec:evidence}

\begin{figure*}[t]
  \centering
  \includegraphics[width=.95\linewidth]{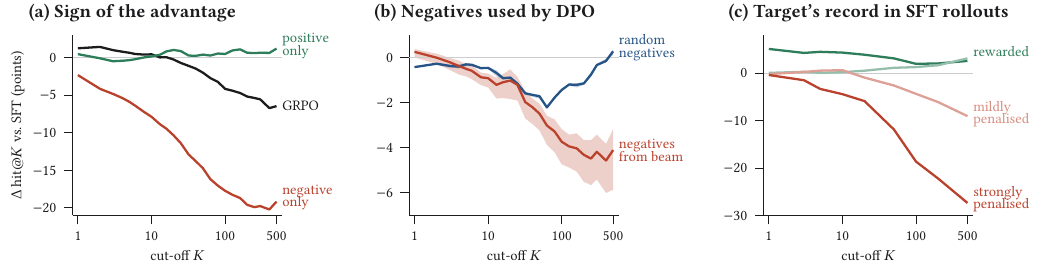}
  \caption{Evidence for the cause (Office; change in \hit{$K$} against SFT; a, c: synced reference, one run).
  (a)~Keeping only the positive advantages removes the loss and the gain; keeping only the negative ones produces a much larger loss.
  (b)~DPO loses recall up to $K=500$ with negatives from the SFT beam and only at intermediate $K$ with random ones (three seeds, bands: range).
  (c)~Test prompts grouped by the rollout record of their target, computed before RL: the loss falls on targets with a negative record.}
  \Description{Three line charts of the change in hit at K against K. Panel a: with positive advantages only the curve stays near zero, GRPO falls to minus 6 points and negative advantages only fall to minus 20 points at K of 500. Panel b: DPO with beam negatives falls to minus 4 points on average, with random negatives it returns to zero at K of 500. Panel c: targets with a strongly negative rollout record lose 27 points at K of 500, mildly negative ones 9 points, and targets with a positive record gain a few points.}
  \label{fig:evidence}
\end{figure*}

The account makes three claims that can be tested: the loss comes from the negative advantages, it depends on which items they fall on, and it reaches the users whose target is penalised (Figure~\ref{fig:evidence}).

First, we zero either sign of the advantages and keep everything else (panel~a).
With positive advantages only, the recall loss is gone on both datasets (\hit{500} changes by $+1.23$ points on Office), and so is the gain at the top: \ndcg{} no longer differs significantly from SFT.
With negative advantages only, \hit{500} falls by 19 to 21 points, three times the loss of the unmodified run on Office and more than ten times on Industrial.
The negative advantages are thus necessary for the loss and, alone, sufficient for a much larger one; the gain needs them too, because the target rises when its competitors are pushed down.

Second, we change which items are penalised (panel~b).
DPO~\citep{rafailov2023dpo} has no rollouts, no group normalisation and no gating; it contrasts the target with one negative per prompt.
With the negative drawn from the SFT model's beam for that prompt, DPO loses 4.1 points of \hit{500} on Office (mean of three seeds), without any gain at the top; with the negative drawn uniformly from the catalogue, \hit{500} is unchanged ($+0.3$) and a loss of up to 2.2 points remains only at intermediate cut-offs.
The damage at large $K$ thus comes from penalising the model's own plausible candidates, and it is not specific to GRPO.

Third, we test whether the loss falls on the users whose target is penalised (panel~c).
The rollout record of an item (Section~\ref{sec:update}) is computed before RL, from the SFT model alone, and is negative when the item would be penalised more than it is rewarded.
We group the test prompts by the record of their target, splitting those with a negative record and those with a positive one at the median, and measure the change in \hit{500} after RL in each group.
On Office the group with the most negative records loses 27.4 points (13.5 on Industrial) and the other negative group 9.0 points, whereas the prompts with a positive record gain about 3 points; the gain at the top is concentrated in the most positive group (\hit{1} $+5.1$).
The few targets that never appear in a training rollout gain rank as well (\hit{500} $+9.6$ points on Office and $+14.7$ on Industrial): items that are not penalised move up as the penalised ones move down.

\subsection{Why this arises in recommendation}\label{sec:whyrec}
GRPO was developed for reasoning tasks with verifiable answers, where lowering a sampled answer that misses the label is justified: the answer is wrong, it is rarely the correct answer to a similar prompt, and only the first answer is used.
Next-item recommendation differs on each count.
The label is not unique: a history is compatible with several next items, of which the log records one, so a penalised candidate is often an acceptable recommendation.
The answers are shared: all users choose from one small closed catalogue, and users with similar histories make similar choices, so the same item is a candidate for some users and the logged item for others.
And the list is used in depth: a retrieval channel consumes ranks up to $10^3$.
\section{Out-of-Rollout Anchoring}\label{sec:method}

\begin{figure}[t]
  \centering
  \includegraphics[width=.95\linewidth]{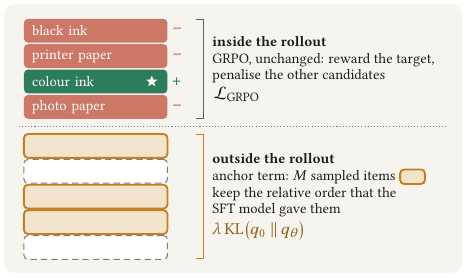}
  \caption{Out-of-rollout anchoring on one training prompt.
  GRPO handles the rollout as before.
  In addition, $M$ items sampled outside the rollout are scored by the frozen SFT model and by the policy, and the policy is penalised if their relative scores change.}
  \Description{A rollout of four candidates, the target with a plus sign and three others with minus signs, labelled inside the rollout, GRPO unchanged. Below a dotted line, five items outside the rollout, three of them highlighted as anchor items that must keep the relative order the SFT model gave them, labelled with the anchor term lambda times a KL divergence.}
  \label{fig:anchor}
\end{figure}

Section~\ref{sec:causes} shows where the damage is done: outside the rollout, on items that lose rank for users whose rollouts do not contain them.
The KL term of GRPO does not reach these items: it is estimated on the rollout too, and so constrains the policy only on its top candidates.
What is needed is a constraint on the items outside the rollout that leaves the update inside it untouched.
To answer RQ3 we therefore propose out-of-rollout anchoring, which is simple and, as Section~\ref{sec:experiments} shows, effective: GRPO is left unchanged, and one added term keeps the order that the SFT model gave to a few items sampled outside the rollout (Figure~\ref{fig:anchor}).
The anchor is also a test of the account: if the damage is done outside the rollout, this one term should restore recall and leave the gain untouched.

\subsection{The anchor term}
For a prompt $x$ with rollout $\cR(x)$, we draw an \emph{anchor set} $\cA(x)\subset\cI\setminus\cR(x)$ of $M$ items and renormalise the SFT model and the policy over it,
\begin{equation}
  q_0(i)=\frac{\exp s_{0}(i\mid x)}{\sum_{i'\in\cA(x)}\exp s_{0}(i'\mid x)},\;\;
  q_\theta(i)=\frac{\exp s_\theta(i\mid x)}{\sum_{i'\in\cA(x)}\exp s_\theta(i'\mid x)},
  \label{eq:renorm}
\end{equation}
for $i\in\cA(x)$, where $s_0$ is the score of Eq.~\eqref{eq:score} under a frozen copy of $\pisft$.
The training loss adds their divergence to the GRPO loss:
\begin{equation}
  \mathcal{L}(\theta;x)=\mathcal{L}_{\mathrm{GRPO}}(\theta;x)+\lambda\,\KL\bigl(q_0\,\|\,q_\theta\bigr).
  \label{eq:anchor}
\end{equation}
The divergence runs from the SFT model to the policy, the direction that penalises lowering items the SFT model supports; it vanishes only if the anchor items keep their SFT score differences, hence their order.
The anchor set is drawn anew at every step, and not at random from the catalogue: it follows the structure of the SIDs.
Items with similar content share their leading codes (Section~\ref{sec:background}), so the items that share leading codes with a rollout candidate are close neighbours of what the policy proposes, which is where the mid-ranked items that GRPO demotes are concentrated (Section~\ref{sec:phenomenon}).
Half of the anchor items are therefore sampled from the items that share the first two codes with a rollout candidate and the other half from those that share the first code.
This neighbourhood is small and well targeted: on the test sets, the items that share a first code with one of the SFT model's top 16 candidates make up 12\% (Office) and 11\% (Industrial) of the catalogue but contain 44\% and 57\% of the targets that SFT ranks between 17 and 500.
We use $M=16$ and $\lambda=0.1$ in all experiments.

\begin{figure*}[t]
  \centering
  \includegraphics[width=\linewidth]{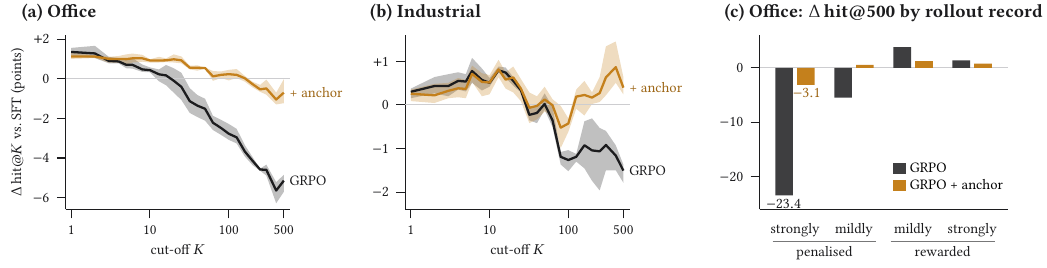}
  \caption{GRPO with and without the anchor.
  (a,~b)~Change in \hit{$K$} against SFT (mean of three seeds, bands show their range).
  (c)~Office test prompts grouped by the rollout record of their target (Section~\ref{sec:evidence}): change in \hit{500} against SFT (one seed).}
  \Description{Three panels. Panels a and b are line charts of the change in hit at K against K for Office and Industrial. The GRPO curve starts above zero and falls to minus 5.1 points on Office and minus 1.5 on Industrial at K of 500; the curve of GRPO with the anchor starts at the same height and stays near zero, ending at minus 0.7 on Office and plus 0.4 on Industrial. Panel c is a bar chart of the change in hit at 500 for four groups of Office test prompts: for targets with a strongly negative rollout record, minus 23.4 points under GRPO and minus 3.1 with the anchor; for mildly negative ones, minus 5.4 and plus 0.6; for the two groups with a positive record, small gains under both.}
  \label{fig:results}
\end{figure*}

\begin{table*}[t]
  \caption{SFT, GRPO and GRPO with the out-of-rollout anchor (mean of three seeds; best value per dataset and column in bold).
  Anchor effect: paired difference between the two RL models in points.}
  \label{tab:main}
  \small\setlength{\tabcolsep}{6.5pt}\renewcommand{\arraystretch}{0.93}
\begin{tabular}{@{}l@{\hspace{8pt}}l@{\hspace{8pt}}rrrrrrrr@{\hspace{14pt}}r@{}}
\toprule
 & & \multicolumn{8}{c}{hit@$K$ (\%)} & NDCG \\
\cmidrule(lr){3-10}
 & & 3 & 5 & 10 & 50 & 100 & 200 & 300 & 500 & @10 \\
\midrule
\multirow{4}{*}{Office} & SFT & 11.76 & 13.42 & 15.97 & 25.73 & 31.94 & \textbf{39.56} & \textbf{44.88} & \textbf{52.84} & 11.59 \\
 & GRPO & 12.66 & 14.13 & 16.41 & 24.22 & 29.16 & 35.42 & 40.29 & 47.69 & 12.55 \\
 & GRPO + anchor & \textbf{12.76} & \textbf{14.41} & \textbf{16.89} & \textbf{26.27} & \textbf{32.17} & 39.29 & 44.40 & 52.14 & \textbf{12.64} \\
 & \textit{anchor effect} & ${+}0.10$ & ${+}0.29$ & ${+}0.48$ & ${+}2.05$ & ${+}3.01$ & ${+}3.87$ & ${+}4.11$ & ${+}4.45$ & ${+}0.09$ \\
\midrule
\multirow{4}{*}{Industrial} & SFT & 8.54 & 10.21 & 13.19 & 23.21 & \textbf{30.20} & 38.05 & 43.97 & 53.23 & 9.05 \\
 & GRPO & \textbf{8.97} & \textbf{10.75} & 13.70 & 23.23 & 28.94 & 37.01 & 43.05 & 51.72 & \textbf{9.48} \\
 & GRPO + anchor & 8.86 & 10.58 & \textbf{13.71} & \textbf{23.33} & 29.77 & \textbf{38.22} & \textbf{44.60} & \textbf{53.63} & 9.42 \\
 & \textit{anchor effect} & ${-}0.11$ & ${-}0.17$ & ${+}0.01$ & ${+}0.10$ & ${+}0.84$ & ${+}1.21$ & ${+}1.55$ & ${+}1.91$ & ${-}0.07$ \\
\midrule
\multirow{4}{*}{Beauty} & SFT & 2.20 & 3.08 & 4.80 & 12.52 & 18.38 & 25.42 & \textbf{30.70} & 37.16 & 2.48 \\
 & GRPO & \textbf{2.45} & 3.36 & 5.27 & 12.85 & 18.07 & 25.03 & 29.89 & 36.67 & 2.86 \\
 & GRPO + anchor & \textbf{2.45} & \textbf{3.55} & \textbf{5.48} & \textbf{13.33} & \textbf{18.45} & \textbf{25.72} & 30.43 & \textbf{37.39} & \textbf{2.95} \\
 & \textit{anchor effect} & ${+}0.00$ & ${+}0.19$ & ${+}0.21$ & ${+}0.47$ & ${+}0.39$ & ${+}0.69$ & ${+}0.53$ & ${+}0.72$ & ${+}0.09$ \\
\midrule
\multirow{4}{*}{Toys} & SFT & 2.36 & 3.46 & 5.00 & 12.32 & 18.12 & 24.98 & 29.36 & 35.18 & 2.72 \\
 & GRPO & \textbf{2.66} & \textbf{3.76} & \textbf{5.50} & 12.56 & 17.72 & 24.36 & 29.26 & 35.14 & \textbf{3.12} \\
 & GRPO + anchor & 2.60 & \textbf{3.76} & \textbf{5.50} & \textbf{13.00} & \textbf{18.82} & \textbf{25.58} & \textbf{29.56} & \textbf{35.60} & 3.03 \\
 & \textit{anchor effect} & ${-}0.06$ & ${+}0.00$ & ${+}0.00$ & ${+}0.44$ & ${+}1.10$ & ${+}1.22$ & ${+}0.30$ & ${+}0.46$ & ${-}0.08$ \\
\midrule
\multirow{4}{*}{Sports} & SFT & 1.54 & 2.14 & 3.70 & 9.24 & 13.20 & 18.86 & 23.26 & 29.88 & 1.90 \\
 & GRPO & \textbf{1.64} & 2.34 & 3.52 & 8.77 & 12.95 & 18.67 & 22.89 & 29.21 & 1.91 \\
 & GRPO + anchor & 1.56 & \textbf{2.39} & \textbf{3.79} & \textbf{9.46} & \textbf{13.53} & \textbf{19.27} & \textbf{23.75} & \textbf{30.11} & \textbf{1.95} \\
 & \textit{anchor effect} & ${-}0.08$ & ${+}0.05$ & ${+}0.27$ & ${+}0.69$ & ${+}0.57$ & ${+}0.59$ & ${+}0.87$ & ${+}0.89$ & ${+}0.04$ \\
\bottomrule
\end{tabular}
\end{table*}

\subsection{Properties}
\emph{The term starts at zero.}
At the SFT initialisation $q_\theta=q_0$, so the term and its gradient vanish.
\emph{It does not resist concentration.}
Both distributions are renormalised over $\cA(x)$, so adding the same constant to the scores of all anchor items leaves the term unchanged: RL can move probability from the tail as a whole to the head at no cost, and only changes in the relative scores of out-of-rollout items are penalised.
\emph{It acts where GRPO is silent}, on every prompt, including the majority whose rollout misses the target and yields no policy gradient.
\emph{It confines the penalty to the histories that drew it.}
The same item is inside the rollout of some users, where GRPO lowers it, and outside that of others, where the anchor asks that its position among the out-of-rollout items stay as the SFT model had it; to satisfy both, the model has to keep the penalty to histories like those that sampled the item (Figure~\ref{fig:results}c).

The term places no constraint on the order among rollout candidates or on absolute ranks.
\section{Experiments}\label{sec:experiments}

\subsection{Setup}
The main comparison is between GRPO and GRPO with the anchor, on the five datasets with three seeds each.
The two arms share the SFT checkpoint and every setting of GRPO.

\subsection{Main results}
\paragraph{Recall is recovered}
The recovery starts as early as the loss (Table~\ref{tab:main}, Figure~\ref{fig:results}a,~b): at $K=50$ the anchored model has the highest \hit{$K$} of the three on all five datasets.
On Office the anchor raises \hit{$K$} at every cut-off from 50 to 500, by 2.1 to 4.4 points and significantly in every seed.
At $K=500$ it reaches 52.1\%, against 47.7\% for GRPO and 52.8\% for SFT: the anchor recovers 4.4 of the 5.1 points lost.
On Industrial the effect grows from 0.8 points at $K=100$ to 1.9 at $K=500$, where it is significant in every seed, and \hit{$K$} exceeds that of the SFT model for $K\ge 200$.
On Beauty, Toys and Sports, where GRPO lost less, the anchor adds 0.3 to 1.2 points at $K\ge 100$ and takes \hit{500} slightly above the SFT model.

\paragraph{The gain at the top is kept}
\ndcg{} of the anchored model does not differ significantly from that of GRPO in any seed on any of the five datasets.
The same is true of \hit{$K$} at $K=5$ and $K=10$, and at $K=10$ the anchored model is level with GRPO or above it on all five datasets, by 0.2 to 0.5 points on Office, Beauty and Sports, possibly because the anchor also brings back targets that GRPO had pushed out of the first ten (Table~\ref{tab:main}).
In Figure~\ref{fig:results}(a,~b) the anchored curve follows GRPO at the top and stays near zero where GRPO turns negative.

\subsection{Where the anchor acts}\label{sec:where}
Three measurements show that the anchor acts outside the rollout and protects the items that GRPO penalises.

\paragraph{The order outside the rollout}
For each test prompt we take the model's 16 highest-scoring items as a stand-in for the rollout and measure the divergence of the anchor term (Eq.~\ref{eq:anchor}) on these 16 items and on 16 anchor items drawn outside them as in training.
Under GRPO the relative scores change more outside the rollout than inside it (1.53 against 1.15 nats on Office; one seed), although all the reward is earned inside.
The anchor cuts the change outside by a factor of eight (to 0.19) and, through the shared network, halves the change inside (to 0.63), at no cost in \ndcg{}; on Industrial they are 0.71 and 0.64 without the anchor, 0.17 and 0.31 with it.

\paragraph{The retrieved set}
Of the test targets that the SFT model ranks in its top 500, GRPO drops 14 to 22\% depending on the dataset (Section~\ref{sec:phenomenon}); the anchored model drops 5 to 8\%.

\paragraph{The penalised targets}
Figure~\ref{fig:results}c repeats the grouping of Figure~\ref{fig:evidence}c for the default configuration.
Among the Office test prompts whose target has the most negative rollout record, \hit{500} falls by 23.4 points under GRPO and by 3.1 with the anchor (10.9 and 2.3 on Industrial).
The prompts whose target has the most positive record, where the gain at the top is made, keep most of it.

\subsection{Other base configurations}\label{sec:pairs}

\begin{table}[t]
  \caption{The anchor on other base configurations: \hit{$K$} in \% (one seed).
  Anchor effect: paired difference in points; a star marks a significant difference.
  Within each pair the better value is bold when the difference is significant.}
  \label{tab:pairs}
  \footnotesize
\begin{tabular}{@{}l@{\hspace{7pt}}rrr@{\hspace{12pt}}rrr@{\hspace{4pt}}}
\toprule
 & \multicolumn{3}{c}{Office} & \multicolumn{3}{c}{Industrial} \\
\cmidrule(lr){2-4}\cmidrule(l){5-7}
 & @50 & @100 & @500 & @50 & @100 & @500 \\
\midrule
SFT & 25.73 & 31.94 & 52.84 & 23.21 & 30.20 & 53.23 \\
\addlinespace[2pt]
Synced reference & 23.76 & 27.81 & 46.40 & 24.22 & 29.36 & 51.71 \\
\quad + anchor & \textbf{25.94} & \textbf{31.28} & \textbf{52.28} & 23.58 & 29.38 & \textbf{53.98} \\
\quad \textit{anchor effect} & ${+}2.18$\rlap{$^{*}$} & ${+}3.47$\rlap{$^{*}$} & ${+}5.88$\rlap{$^{*}$} & ${-}0.64$ & ${+}0.02$ & ${+}2.27$\rlap{$^{*}$} \\
\addlinespace[2pt]
REINFORCE++ & 20.86 & 24.39 & 37.92 & 21.51 & 26.34 & 46.94 \\
\quad + anchor & \textbf{25.46} & \textbf{30.99} & \textbf{51.17} & \textbf{22.61} & \textbf{28.72} & \textbf{52.42} \\
\quad \textit{anchor effect} & ${+}4.60$\rlap{$^{*}$} & ${+}6.60$\rlap{$^{*}$} & ${+}13.26$\rlap{$^{*}$} & ${+}1.10$\rlap{$^{*}$} & ${+}2.38$\rlap{$^{*}$} & ${+}5.47$\rlap{$^{*}$} \\
\bottomrule
\end{tabular}
\end{table}

Table~\ref{tab:pairs} adds the anchor to two other base configurations: GRPO with a synced reference and REINFORCE++ (Section~\ref{sec:robust}).
In every pair the anchor raises \hit{500} significantly.
With REINFORCE++ the anchor raises \hit{500} by 13.3 points on Office and 5.5 on Industrial, and \ndcg{} rises significantly as well (from 11.45 to 12.24 and from 8.96 to 9.32).
\section{Related Work}\label{sec:related}

\paragraph{Generative recommendation and its RL post-training}
Generative recommendation with semantic IDs~\citep{rajput2023tiger,decao2021genre,tay2022dsi,wang2024letter,wang2024eager,geng2022p5,bao2023tallrec,zheng2024lcrec} adopted RL through the OneRec systems~\citep{deng2025onerec,zhou2025onerectr,zhou2025onerecv2} and MiniOneRec~\citep{kong2025minionerec}; OneRec-Think, OpenOneRec, Rec-R1 and Rank-GRPO use the same family of objectives~\citep{liu2025onerecthink,zhou2025openonerec,lin2025recr1,zhu2025rankgrpo}, alongside DPO-style ones~\citep{chen2024sdpo,gao2025sprec}.
A recent line of work observes that most rollouts contain no target and therefore no signal, and repairs this with modified groups, target-prefix hints, prefix-level credit or step-aligned rewards~\citep{zhang2026recast,zhang2026hcgrec,yu2026daso,zheng2026sapo,chen2026lasar,jiang2026vstar,xie2026sage}.
Evaluated at $K\le 10$, in a few cases up to 32, these methods do not see what we find: the rollouts that do carry signal are the source of a recall loss.

\paragraph{Reports of recall loss under post-training}
Several other industrial papers report observations that agree with ours.
GenRec finds that GRPO without its relevance gate lowers HR@50 below the SFT model~\citep{zou2026genrec}, and RGD reports a lower HitRate@604 after DPO or GRPO alignment although the reward rises~\citep{yang2026rgd}.
In PinGER and Multi-Decoder OneRec recall collapses unless an additional term ties the model to the pretrained one~\citep{li2026pinger,wang2026multidecoder}, and GrowthGR describes a trade-off between Recall@10 and Recall@1000 that depends on the rollout strategy~\citep{wang2026growthgr}.
TSGR notes that rollouts concentrate on head candidates, so that reward normalisation suppresses items still within the top 1{,}000~\citep{zhan2026tsgr}.
Each is a single observation; we provide a controlled measurement across $K$ and an explanation that accounts for the gain and the loss together.

\paragraph{RL post-training of language models and hard negatives}
In reasoning tasks, RL with verifiable rewards improves pass@1 while base models keep a higher pass@$k$ at large $k$~\citep{yue2025rlvr}; that effect concerns the coverage of sampling, whereas ours changes the deterministic ranking of a closed item set and transfers across users.
Existing regularisers towards the initial model~\citep{schulman2017ppo,shao2024deepseekmath,ouyang2022instructgpt,gorbatovski2024trdpo,li2026pinger,wang2026multidecoder} act on the samples that the policy or the data provide, while our anchor is restricted to items outside the rollout and to their relative scores.
The negatives of GRPO in recommendation are hard negatives in the sense of dense retrieval and implicit-feedback recommendation, where candidates mined from a model's own top results are often unlabelled positives~\citep{xiong2021ance,qu2021rocketqa,rendle2009bpr,ding2020srns}.
\section{Conclusion and Future Work}\label{sec:conclusion}

RL post-training improves a generative recommender where it is usually evaluated, at the top of the list, and damages it where it is usually deployed, in the candidate list that it hands on.
On all five datasets the same model becomes a better ranker and a worse retriever, and an evaluation confined to the top of the list sees only the gain.
The root cause is that only one next item is logged per user: RL penalises the model's other candidates as if they were wrong, although they are its most plausible alternatives and often the next items of similar users, and because the model generalises across users, these items drop out of the candidate lists of the users who would have chosen them.
The remedy is as simple as the cause.
Out-of-rollout anchoring adds a single term that keeps the SFT model's order of the items outside the rollout; it preserves the gain at the top, restores recall to the level of the SFT model, and helps under every RL configuration we tried.
We regard the diagnosis as the main result: it explains why the anchor suffices and where a better reward should act.
We recommend that RL post-training be evaluated at the depth at which the model is deployed, not only at the top of the list.

These results open three directions.
The anchor preserves recall and often lifts it slightly above the SFT level.
RL itself still has no learning signal for targets outside the rollout; an RL objective designed for large $K$, with rollouts and rewards that reach deep into the list, could strengthen recall itself and improve both ends of the list at once.
The root cause can also be addressed at the reward: one that recognises several acceptable items per user, for example the choices of users with similar histories, would not penalise plausible candidates in the first place.
Finally, the trade-off should be measured at industrial scale, with larger catalogues and models and with the downstream ranker in the loop, where the recall of the candidate list becomes an online gain.

\bibliographystyle{ACM-Reference-Format}
\bibliography{refs}

@inproceedings{bao2023tallrec,
  author    = {Keqin Bao and Jizhi Zhang and Yang Zhang and Wenjie Wang and Fuli Feng and Xiangnan He},
  title     = {{TALLRec: An Effective and Efficient Tuning Framework to Align Large Language Model with Recommendation}},
  booktitle = {Proceedings of the ACM Conference on Recommender Systems (RecSys)},
  year      = {2023},
  note      = {arXiv:2305.00447}
}

@inproceedings{chen2024sdpo,
  author    = {Yuxin Chen and Junfei Tan and An Zhang and Zhengyi Yang and Leheng Sheng and Enzhi Zhang and Xiang Wang and Tat-Seng Chua},
  title     = {{On Softmax Direct Preference Optimization for Recommendation}},
  booktitle = {Advances in Neural Information Processing Systems (NeurIPS)},
  year      = {2024},
  note      = {arXiv:2406.09215}
}

@article{chen2025onesearch,
  author  = {Ben Chen and Xian Guo and Siyuan Wang and Zihan Liang and Yue Lv and Yufei Ma and Xinlong Xiao and Bowen Xue and others},
  title   = {{OneSearch: A Preliminary Exploration of the Unified End-to-End Generative Framework for E-commerce Search}},
  journal = {arXiv preprint arXiv:2509.03236},
  year    = {2025}
}

@article{chen2026lasar,
  author  = {Yiwen Chen and Fuwei Zhang and Zehao Chen and Hehan Li and Peizhi Xu and Hanmeng Liu and Shuanglong Li and Xin Pei and others},
  title   = {{LASAR: Latent Adaptive Semantic Aligned Reasoning for Generative Recommendation}},
  journal = {arXiv preprint arXiv:2605.10207},
  year    = {2026}
}

@article{chen2026onesearchv2,
  author  = {Ben Chen and Siyuan Wang and Yufei Ma and Zihan Liang and Xuxin Zhang and Yue Lv and Ying Yang and Huangyu Dai and others},
  title   = {{OneSearch-V2: The Latent Reasoning Enhanced Self-distillation Generative Search Framework}},
  journal = {arXiv preprint arXiv:2603.24422},
  year    = {2026}
}

@inproceedings{decao2021genre,
  author    = {Nicola De Cao and Gautier Izacard and Sebastian Riedel and Fabio Petroni},
  title     = {{Autoregressive Entity Retrieval}},
  booktitle = {International Conference on Learning Representations (ICLR)},
  year      = {2021},
  note      = {arXiv:2010.00904}
}

@article{deng2025onerec,
  author  = {Jiaxin Deng and Shiyao Wang and Kuo Cai and Lejian Ren and Qigen Hu and Weifeng Ding and Qiang Luo and Guorui Zhou},
  title   = {{OneRec: Unifying Retrieve and Rank with Generative Recommender and Iterative Preference Alignment}},
  journal = {arXiv preprint arXiv:2502.18965},
  year    = {2025}
}

@inproceedings{ding2020srns,
  author    = {Jingtao Ding and Yuhan Quan and Quanming Yao and Yong Li and Depeng Jin},
  title     = {{Simplify and Robustify Negative Sampling for Implicit Collaborative Filtering}},
  booktitle = {Advances in Neural Information Processing Systems (NeurIPS)},
  year      = {2020},
  note      = {arXiv:2009.03376}
}

@inproceedings{gao2025sprec,
  author    = {Chongming Gao and Ruijun Chen and Shuai Yuan and Kexin Huang and Yuanqing Yu and Xiangnan He},
  title     = {{SPRec: Self-Play to Debias LLM-based Recommendation}},
  booktitle = {Proceedings of the ACM Web Conference (WWW)},
  year      = {2025},
  note      = {arXiv:2412.09243}
}

@inproceedings{geng2022p5,
  author    = {Shijie Geng and Shuchang Liu and Zuohui Fu and Yingqiang Ge and Yongfeng Zhang},
  title     = {{Recommendation as Language Processing (RLP): A Unified Pretrain, Personalized Prompt \& Predict Paradigm (P5)}},
  booktitle = {Proceedings of the ACM Conference on Recommender Systems (RecSys)},
  year      = {2022},
  note      = {arXiv:2203.13366}
}

@article{gorbatovski2024trdpo,
  author  = {Alexey Gorbatovski and Boris Shaposhnikov and Alexey Malakhov and Nikita Surnachev and Yaroslav Aksenov and Ian Maksimov and Nikita Balagansky and Daniil Gavrilov},
  title   = {{Learn Your Reference Model for Real Good Alignment}},
  journal = {arXiv preprint arXiv:2404.09656},
  year    = {2024}
}

@inproceedings{he2016ups,
  author    = {Ruining He and Julian McAuley},
  title     = {{Ups and Downs: Modeling the Visual Evolution of Fashion Trends with One-Class Collaborative Filtering}},
  booktitle = {Proceedings of the International Conference on World Wide Web (WWW)},
  year      = {2016},
  note      = {arXiv:1602.01585}
}

@inproceedings{he2025plum,
  author    = {Ruining He and Lukasz Heldt and Lichan Hong and Raghunandan Keshavan and Shifan Mao and Nikhil Mehta and Zhengyang Su and Alicia Tsai and others},
  title     = {{PLUM: Adapting Pre-trained Language Models for Industrial-scale Generative Recommendations}},
  booktitle = {Proceedings of the ACM Web Conference (WWW)},
  year      = {2026},
  note      = {arXiv:2510.07784}
}

@article{hu2025reinforcepp,
  author  = {Jian Hu and Jason Klein Liu and Haotian Xu and Wei Shen},
  title   = {{REINFORCE++: Stabilizing Critic-Free Policy Optimization with Global Advantage Normalization}},
  journal = {arXiv preprint arXiv:2501.03262},
  year    = {2025}
}

@article{jiang2026vstar,
  author  = {Jie Jiang and Yangru Huang and Zeyu Wang and Changping Wang and Yuling Xiong and Jun Zhang and Huan Yu},
  title   = {{Spend Search Where It Pays: Value-Guided Structured Sampling and Optimization for Generative Recommendation}},
  journal = {arXiv preprint arXiv:2602.10699},
  year    = {2026}
}

@inproceedings{kang2018sasrec,
  author    = {Wang-Cheng Kang and Julian McAuley},
  title     = {{Self-Attentive Sequential Recommendation}},
  booktitle = {IEEE International Conference on Data Mining (ICDM)},
  year      = {2018},
  note      = {arXiv:1808.09781}
}

@article{kong2025minionerec,
  author  = {Xiaoyu Kong and Leheng Sheng and Junfei Tan and Yuxin Chen and Jiancan Wu and An Zhang and Xiang Wang and Xiangnan He},
  title   = {{MiniOneRec: An Open-Source Framework for Scaling Generative Recommendation}},
  journal = {arXiv preprint arXiv:2510.24431},
  year    = {2025}
}

@inproceedings{korbak2022rlkl,
  author    = {Tomasz Korbak and Ethan Perez and Christopher L Buckley},
  title     = {{RL with KL penalties is better viewed as Bayesian inference}},
  booktitle = {Findings of the Association for Computational Linguistics: EMNLP},
  year      = {2022},
  note      = {arXiv:2205.11275}
}

@inproceedings{lee2022rqvae,
  author    = {Doyup Lee and Chiheon Kim and Saehoon Kim and Minsu Cho and Wook-Shin Han},
  title     = {{Autoregressive Image Generation using Residual Quantization}},
  booktitle = {IEEE/CVF Conference on Computer Vision and Pattern Recognition (CVPR)},
  year      = {2022},
  note      = {arXiv:2203.01941}
}

@article{li2024genrpo,
  author  = {Mingming Li and Huimu Wang and Zuxu Chen and Guangtao Nie and Yiming Qiu and Guoyu Tang and Lin Liu and Jingwei Zhuo},
  title   = {{Generative Retrieval with Preference Optimization for E-commerce Search}},
  journal = {arXiv preprint arXiv:2407.19829},
  year    = {2024}
}

@article{li2026pinger,
  author  = {Xinyi Li and Sujan Reddy Ale and Yi-Ping Hsu and Zelun Wang and Yilin Chen and Hanyu Li and Hanlin Lu and Arkin Dharawat and others},
  title   = {{PinGER: Token-Efficient Generative Retrieval with Semantic IDs}},
  journal = {arXiv preprint arXiv:2610.33435},
  year    = {2026}
}

@article{lin2025recr1,
  author    = {Jiacheng Lin and Tian Wang and Kun Qian},
  title     = {{Rec-R1: Bridging Generative Large Language Models and User-Centric Recommendation Systems via Reinforcement Learning}},
  journal   = {Transactions on Machine Learning Research (TMLR)},
  year      = {2025},
  note      = {arXiv:2503.24289}
}

@article{liu2025onerecthink,
  author  = {Zhanyu Liu and Shiyao Wang and Xingmei Wang and Rongzhou Zhang and Jiaxin Deng and Honghui Bao and Jinghao Zhang and Wuchao Li and others},
  title   = {{OneRec-Think: In-Text Reasoning for Generative Recommendation}},
  journal = {arXiv preprint arXiv:2510.11639},
  year    = {2025}
}

@inproceedings{mcauley2015image,
  author    = {Julian McAuley and Christopher Targett and Qinfeng Shi and Anton van den Hengel},
  title     = {{Image-based Recommendations on Styles and Substitutes}},
  booktitle = {Proceedings of the International ACM SIGIR Conference on Research and Development in Information Retrieval (SIGIR)},
  year      = {2015},
  note      = {arXiv:1506.04757}
}

@article{minimax2025m1,
  author  = {{MiniMax}},
  title   = {{MiniMax-M1: Scaling Test-Time Compute Efficiently with Lightning Attention}},
  journal = {arXiv preprint arXiv:2506.13585},
  year    = {2025}
}

@inproceedings{ouyang2022instructgpt,
  author    = {Long Ouyang and Jeff Wu and Xu Jiang and Diogo Almeida and Carroll L. Wainwright and Pamela Mishkin and Chong Zhang and Sandhini Agarwal and others},
  title     = {{Training language models to follow instructions with human feedback}},
  booktitle = {Advances in Neural Information Processing Systems (NeurIPS)},
  year      = {2022},
  note      = {arXiv:2203.02155}
}

@inproceedings{qu2021rocketqa,
  author    = {Yingqi Qu and Yuchen Ding and Jing Liu and Kai Liu and Ruiyang Ren and Wayne Xin Zhao and Daxiang Dong and Hua Wu and others},
  title     = {{RocketQA: An Optimized Training Approach to Dense Passage Retrieval for Open-Domain Question Answering}},
  booktitle = {Proceedings of the Conference of the North American Chapter of the Association for Computational Linguistics (NAACL)},
  year      = {2021},
  note      = {arXiv:2010.08191}
}

@article{qwen2024qwen25,
  author  = {{Qwen Team}},
  title   = {{Qwen2.5 Technical Report}},
  journal = {arXiv preprint arXiv:2412.15115},
  year    = {2024}
}

@inproceedings{rafailov2023dpo,
  author    = {Rafael Rafailov and Archit Sharma and Eric Mitchell and Stefano Ermon and Christopher D. Manning and Chelsea Finn},
  title     = {{Direct Preference Optimization: Your Language Model is Secretly a Reward Model}},
  booktitle = {Advances in Neural Information Processing Systems (NeurIPS)},
  year      = {2023},
  note      = {arXiv:2305.18290}
}

@inproceedings{rajput2023tiger,
  author    = {Shashank Rajput and Nikhil Mehta and Anima Singh and Raghunandan H. Keshavan and Trung Vu and Lukasz Heldt and Lichan Hong and Yi Tay and others},
  title     = {{Recommender Systems with Generative Retrieval}},
  booktitle = {Advances in Neural Information Processing Systems (NeurIPS)},
  year      = {2023},
  note      = {arXiv:2305.05065}
}

@inproceedings{rendle2009bpr,
  author    = {Steffen Rendle and Christoph Freudenthaler and Zeno Gantner and Lars Schmidt-Thieme},
  title     = {{BPR: Bayesian Personalized Ranking from Implicit Feedback}},
  booktitle = {Proceedings of the Conference on Uncertainty in Artificial Intelligence (UAI)},
  year      = {2009},
  note      = {arXiv:1205.2618}
}

@article{schulman2017ppo,
  author  = {John Schulman and Filip Wolski and Prafulla Dhariwal and Alec Radford and Oleg Klimov},
  title   = {{Proximal Policy Optimization Algorithms}},
  journal = {arXiv preprint arXiv:1707.06347},
  year    = {2017}
}

@article{shao2024deepseekmath,
  author  = {Zhihong Shao and Peiyi Wang and Qihao Zhu and Runxin Xu and Junxiao Song and Xiao Bi and Haowei Zhang and Mingchuan Zhang and others},
  title   = {{DeepSeekMath: Pushing the Limits of Mathematical Reasoning in Open Language Models}},
  journal = {arXiv preprint arXiv:2402.03300},
  year    = {2024}
}

@inproceedings{tay2022dsi,
  author    = {Yi Tay and Vinh Q. Tran and Mostafa Dehghani and Jianmo Ni and Dara Bahri and Harsh Mehta and Zhen Qin and Kai Hui and others},
  title     = {{Transformer Memory as a Differentiable Search Index}},
  booktitle = {Advances in Neural Information Processing Systems (NeurIPS)},
  year      = {2022},
  note      = {arXiv:2202.06991}
}

@inproceedings{wang2024eager,
  author    = {Ye Wang and Jiahao Xun and Minjie Hong and Jieming Zhu and Tao Jin and Wang Lin and Haoyuan Li and Linjun Li and others},
  title     = {{EAGER: Two-Stream Generative Recommender with Behavior-Semantic Collaboration}},
  booktitle = {Proceedings of the ACM SIGKDD Conference on Knowledge Discovery and Data Mining (KDD)},
  year      = {2024},
  note      = {arXiv:2406.14017}
}

@inproceedings{wang2024letter,
  author    = {Wenjie Wang and Honghui Bao and Xinyu Lin and Jizhi Zhang and Yongqi Li and Fuli Feng and See-Kiong Ng and Tat-Seng Chua},
  title     = {{Learnable Item Tokenization for Generative Recommendation}},
  booktitle = {Proceedings of the ACM International Conference on Information and Knowledge Management (CIKM)},
  year      = {2024},
  note      = {arXiv:2405.07314}
}

@article{wang2026growthgr,
  author  = {Yifan Wang and Yixuan Wang and YiDan Liang and Qiang Liu and Fei Xiao},
  title   = {{Towards Sustainable Growth: A Multi-Value-Aware Retrieval Framework for E-Commerce Search}},
  journal = {arXiv preprint arXiv:2605.17994},
  year    = {2026}
}

@article{wang2026multidecoder,
  author  = {You Wang and Zhao Liu and Guoping Tang and Yiqing Yang and Shuo Su and Jing Liu and Naifu Zhou and Xiaoyou Zhou and others},
  title   = {{Multi-Decoder OneRec: Controllable Generative Retrieval for Multi-Objective Industrial Recommendation}},
  journal = {arXiv preprint arXiv:2607.26500},
  year    = {2026}
}

@article{xie2026sage,
  author  = {Yu Xie and Xing Kai Ren and Ying Qi and Hu Yao},
  title   = {{SAGE: Sequence-level Adaptive Gradient Evolution for Generative Recommendation}},
  journal = {arXiv preprint arXiv:2601.21452},
  year    = {2026}
}

@inproceedings{xiong2021ance,
  author    = {Lee Xiong and Chenyan Xiong and Ye Li and Kwok-Fung Tang and Jialin Liu and Paul Bennett and Junaid Ahmed and Arnold Overwijk},
  title     = {{Approximate Nearest Neighbor Negative Contrastive Learning for Dense Text Retrieval}},
  booktitle = {International Conference on Learning Representations (ICLR)},
  year      = {2021},
  note      = {arXiv:2007.00808}
}

@article{yang2026rgd,
  author  = {Ruochen Yang and Yusheng Huang and Youfeng Zheng and Shuang Wen and Liangliang Chen and Pengbo Xu and Xiaoyu Zhang and Shijun Wang and others},
  title   = {{Reward Guided Decoding for Generative Recommendation}},
  journal = {arXiv preprint arXiv:2607.25344},
  year    = {2026}
}

@article{yu2025dapo,
  author  = {Qiying Yu and Zheng Zhang and Ruofei Zhu and Yufeng Yuan and Xiaochen Zuo and Yu Yue and Weinan Dai and Tiantian Fan and others},
  title   = {{DAPO: An Open-Source LLM Reinforcement Learning System at Scale}},
  journal = {arXiv preprint arXiv:2503.14476},
  year    = {2025}
}

@article{yu2026daso,
  author  = {Xin Yu and Stephen Li and Sina Aghaei and Zifan Zhu and Jiamu Bai and Guanjie Huang and Bo Peng and Yiyao Liu and others},
  title   = {{Difficulty-Aware Semantic-ID Optimization for Generative Recommendation}},
  journal = {arXiv preprint arXiv:2608.20611},
  year    = {2026}
}

@inproceedings{yue2025rlvr,
  author    = {Yang Yue and Zhiqi Chen and Rui Lu and Andrew Zhao and Zhaokai Wang and Yang Yue and Shiji Song and Gao Huang},
  title     = {{Does Reinforcement Learning Really Incentivize Reasoning Capacity in LLMs Beyond the Base Model?}},
  booktitle = {Advances in Neural Information Processing Systems (NeurIPS)},
  year      = {2025},
  note      = {arXiv:2504.13837}
}

@article{zhan2026tsgr,
  author  = {Tianyu Zhan and Gui Ling and Tong Xiong and Kunhai Lin and Yang Wang and Kaixuan Zhang and Zhihong Chen and Yuliang Yan and others},
  title   = {{TSGR: Taobao Search Generative Retrieval}},
  journal = {arXiv preprint arXiv:2607.18796},
  year    = {2026}
}

@inproceedings{zhang2026hcgrec,
  author    = {Kangning Zhang and Haotian Fang and Xukun Luo and Hao Yin and Yang Gao and Peng Yan and Weiwen Liu and Weinan Zhang and others},
  title     = {{Learning from Unreachable Rewards: Hint-Conditioned Reinforcement Learning for Generative Recommendation}},
  booktitle = {Proceedings of the ACM International Conference on Information and Knowledge Management (CIKM)},
  year      = {2026},
  note      = {arXiv:2608.11980}
}

@article{zhang2026recast,
  author  = {Peiyan Zhang and Hanmo Liu and Chengxuan Tong and Yuxia Wu and Wei Guo and Yong Liu},
  title   = {{ReCast: Recasting Learning Signals for Reinforcement Learning in Generative Recommendation}},
  journal = {arXiv preprint arXiv:2604.22169},
  year    = {2026}
}

@inproceedings{zheng2024lcrec,
  author    = {Bowen Zheng and Yupeng Hou and Hongyu Lu and Yu Chen and Wayne Xin Zhao and Ming Chen and Ji-Rong Wen},
  title     = {{Adapting Large Language Models by Integrating Collaborative Semantics for Recommendation}},
  booktitle = {IEEE International Conference on Data Engineering (ICDE)},
  year      = {2024},
  note      = {arXiv:2311.09049}
}

@article{zheng2025gspo,
  author  = {Chujie Zheng and Shixuan Liu and Mingze Li and Xiong-Hui Chen and Bowen Yu and Chang Gao and Kai Dang and Yuqiong Liu and others},
  title   = {{Group Sequence Policy Optimization}},
  journal = {arXiv preprint arXiv:2507.18071},
  year    = {2025}
}

@article{zheng2026sapo,
  author  = {Zaiyi Zheng and Liang Wu and Guanghui Min and Yaochen Zhu and Liangjie Hong and Chen Chen and Jundong Li},
  title   = {{SAPO: Step-Aligned Policy Optimization for Reasoning-Based Generative Recommendation}},
  journal = {arXiv preprint arXiv:2605.17648},
  year    = {2026}
}

@article{zhou2025onerectr,
  author  = {Guorui Zhou and Jiaxin Deng and Jinghao Zhang and Kuo Cai and Lejian Ren and Qiang Luo and Qianqian Wang and Qigen Hu and others},
  title   = {{OneRec Technical Report}},
  journal = {arXiv preprint arXiv:2506.13695},
  year    = {2025}
}

@article{zhou2025onerecv2,
  author  = {Guorui Zhou and Hengrui Hu and Hongtao Cheng and Huanjie Wang and Jiaxin Deng and Jinghao Zhang and Kuo Cai and Lejian Ren and others},
  title   = {{OneRec-V2 Technical Report}},
  journal = {arXiv preprint arXiv:2508.20900},
  year    = {2025}
}

@article{zhou2025openonerec,
  author  = {Guorui Zhou and Honghui Bao and Jiaming Huang and Jiaxin Deng and Jinghao Zhang and Junda She and Kuo Cai and Lejian Ren and others},
  title   = {{OpenOneRec Technical Report}},
  journal = {arXiv preprint arXiv:2512.24762},
  year    = {2025}
}

@article{zhu2025rankgrpo,
  author  = {Yaochen Zhu and Harald Steck and Dawen Liang and Yinhan He and Vito Ostuni and Jundong Li and Nathan Kallus},
  title   = {{Rank-GRPO: Training LLM-based Conversational Recommender Systems with Reinforcement Learning}},
  journal = {arXiv preprint arXiv:2510.20150},
  year    = {2025}
}

@article{zhu2026eggrpo,
  author  = {Jianbo Zhu and Xing Fang and Jing Wang and Mingmin Jin and Bokang Wang and Guangxin Song and Zhenyu Xie and Junjie Bai},
  title   = {{Efficient Generative Retrieval for E-commerce Search with Semantic Cluster IDs and Expert-Guided RL}},
  journal = {arXiv preprint arXiv:2605.14434},
  year    = {2026}
}

@inproceedings{zou2026genrec,
  author    = {Yanyan Zou and Junbo Qi and Lunsong Huang and Yu Li and Kewei Xu and Jiabao Gao and Binglei Zhao and Xuanhua Yang and others},
  title     = {{GenRec: A Preference-Oriented Generative Framework for Large-Scale Recommendation}},
  booktitle = {Proceedings of the International ACM SIGIR Conference on Research and Development in Information Retrieval (SIGIR)},
  year      = {2026},
  note      = {arXiv:2604.14878}
}

@inproceedings{ni2019justifying,
  author    = {Jianmo Ni and Jiacheng Li and Julian McAuley},
  title     = {{Justifying Recommendations using Distantly-Labeled Reviews and Fine-Grained Aspects}},
  booktitle = {Proceedings of the Conference on Empirical Methods in Natural Language Processing and the International Joint Conference on Natural Language Processing (EMNLP-IJCNLP)},
  pages     = {188--197},
  year      = {2019}
}

@article{jarvelin2002cumulated,
  author  = {Kalervo J{\"a}rvelin and Jaana Kek{\"a}l{\"a}inen},
  title   = {{Cumulated Gain-Based Evaluation of IR Techniques}},
  journal = {ACM Transactions on Information Systems},
  volume  = {20},
  number  = {4},
  pages   = {422--446},
  year    = {2002}
}

@book{efron1994bootstrap,
  author    = {Bradley Efron and Robert J. Tibshirani},
  title     = {{An Introduction to the Bootstrap}},
  publisher = {Chapman \& Hall/CRC},
  year      = {1994}
}

@misc{vonwerra2020trl,
  author       = {Leandro von Werra and Younes Belkada and Lewis Tunstall and Edward Beeching and Tristan Thrush and Nathan Lambert and Shengyi Huang and Kashif Rasul and Quentin Gallou{\'e}dec},
  title        = {{TRL: Transformer Reinforcement Learning}},
  howpublished = {\url{https://github.com/huggingface/trl}},
  year         = {2020}
}

\end{document}